\documentclass[10pt]{article}

\usepackage{amsmath}
\usepackage{amsfonts}
\usepackage{amssymb}
\usepackage{amsthm}
\usepackage{stmaryrd} 
\usepackage{braket}
\usepackage{empheq}
\usepackage{graphicx}
\usepackage{subfigure}
\usepackage{hyperref}
\usepackage{footmisc}
\usepackage{appendix}
\usepackage{epsf}
\usepackage{epsfig}
\usepackage{tikz,tikz-3dplot}
\usetikzlibrary{decorations.markings,decorations.pathmorphing}
\usepackage{caption}
\usepackage{pgfplots}
\usepackage[utf8]{inputenc}
\usepackage[english]{babel}
\usepackage[T1]{fontenc}
\usepackage{sectsty}
\usepackage{authblk}
\subsubsectionfont{\normalfont\itshape}

\usepackage{fullpage}

\renewcommand{\leq}{\leqslant}
\renewcommand{\geq}{\geqslant}

\newcommand{\triplenorm}[1]{|\mskip -2 mu|\mskip -2 mu|#1|\mskip -2mu |\mskip -2mu|}
\newcommand{\bbone}{{\text{\usefont{U}{bbold}{m}{n}\char49}}}

\DeclareMathOperator{\tr}{tr}

\theoremstyle{plain}
\newtheorem{theorem}{Theorem}[section]
\newtheorem{proposition}[theorem]{Proposition}

\newtheorem{lemma}[theorem]{Lemma}

\theoremstyle{definition}
\newtheorem{definition}[theorem]{Definition}
\newtheorem{remark}[theorem]{Remark}

\title{Quantifying quantum coherence \\and the deviation from the total probability formula}
\author[a b]{Antoine Soulas}
\affil[a]{Faculty of Physics, University of Vienna \\ Boltzmanngasse 5, 1090 Vienna (Austria); antoine.soulas@univie.ac.at}
\affil[b]{IQOQI Vienna, Austrian Academy of Sciences, Boltzmanngasse 3, 1090 Vienna (Austria)}
\date{}

\begin{document}
\maketitle

\abstract{We propose a novel approach to quantify quantum coherence which, contrary to the previous ones, does not rely on resource theory but rather on ontological considerations. In this framework, coherence is understood as the ability for a quantum system's statistics to deviate from the total probability formula. After motivating the importance of the total probability formula in quantum foundations, we propose a new set of axioms that a measure of coherence should satisfy, and show that it defines a class of measures different from the main previous proposal. Finally, we prove a general result about the dependence of the $l^2$-coherence norm on the basis of interest, and show that it is well approximated by the square root of the purity in most bases.}


\newpage
\section{Introduction} \label{intro}
There exists a wide variety of measures of coherence. The pioneering work of Åberg \cite{aberg2006quantifying}, followed by the one of Baumgratz, Cramer, and Plenio \cite{baumgratz2014quantifying} have sparkled a strong interest in the task of quantifying coherence from the perspective of resource theory \cite{chitambar2016comparison, winter2016operational, yadin2016general, streltsov2017colloquium, hu2018quantum, chitambar2019quantum, budiyono2023quantifying}. In these studies, a measure of coherence is generally a map associating, to any density matrix, a nonnegative real number representing its level of coherence. In \cite{baumgratz2014quantifying}, the authors have attempted to ‘put such measures on a sound footing by establishing a quantitative theory of coherence as a resource’. To do so, they have identified a property called ‘monotonicity under incoherent operations’ as the key ingredient. However, thinking within the framework of resource theory, as useful as it may be, is not without drawbacks. In particular: (i) a specific basis in which to quantify coherence (the computational basis) is usually selected once and for all \cite{chitambar2016critical}, hence the formalism is not very friendly with the study of how quantum coherence depends on the choice of basis; (ii) a clear physical meaning for these measures is generally lacking \cite{marvian2016quantify}.

In this paper, we would like to pursue a similar aim as in \cite{baumgratz2014quantifying} and propose a new set of axioms that a measure of coherence should satisfy, from an entirely different point of view. Importantly, it will solve the issues (i) and (ii) mentioned above. In our formalism, the level of coherence will now be a function of both the state and the basis in which it is considered, and the axioms will be physically well-motivated. The perspective embraced we be more ontological, \textit{i.e.} focusing on what quantum mechanics tells us about the nature of reality. 

In Section \ref{TPF}, we motivate the philosophical significance of the violation of the total probability formula in quantum mechanics. This will lead us to understand quantum coherence as the ability for a quantum system to deviate from the total probability formula. To formalize this idea mathematically, a proper notion of distance between orthonormal bases will be necessary, as detailed in Section \ref{distance}. Our main results are found in Section \ref{axioms}, where we propose a new set of axioms for a measure of coherence (Definition \ref{definition}) and check whether they are satisfied by several candidates (Theorem \ref{main}). In particular, we show that the $l^1$ and $l^2$-norms of coherence are measures of decoherence, but not the relative entropy of coherence. Finally, Section \ref{l2} presents a result on the $l^2$ norm (Proposition \ref{purity}) illustrating the new possibilities offered by our framework to study the dependance of coherence on the basis of interest.

\paragraph{Notations} Note that all the following discussion is restricted to the finite dimensional case. In the sequel, we will denote:
\begin{itemize}
\item $\mathcal{H}$ a Hilbert space of dimension $n$;
\item $\mathcal{U}(\mathcal{H})$ the set of unitary operators over $\mathcal{H}$;
\item $\mathcal{S}(\mathcal{H})$ the set of density matrices over $\mathcal{H}$;
\item $\mathfrak{B}(\mathcal{H})$ the set of orthonormal bases in $\mathcal{H}$;
\item for $\rho \in \mathcal{S}(\mathcal{H})$, $S(\rho)$ the Von Neumann entropy of $\rho$;
\item $\triplenorm{.}$ the usual operator norm on bounded operators over $\mathcal{H}$;
\item for $\rho \in \mathcal{S}(\mathcal{H})$ and $\mathcal{B} \in \mathfrak{B}(\mathcal{H})$, $D_{\mathcal{B}}[\rho]$ the diagonal part of $\rho$ as written in $\mathcal{B}$ and $Q_{\mathcal{B}} [\rho]$ the remaining non-diagonal part;
\item $\mathcal{B}_{\hat{A}}$ the eigenbasis of a Hermitian operator $\hat{A}$. This definition could be ambiguous, since the eigenbasis may not be unique in the case of a degenerate spectrum. In this paper, \textbf{whenever the notation $\mathcal{B}_{\hat{A}}$ is employed, the statement must be understood as being true for any eigenbasis of $\hat{A}$}.
\end{itemize}

\section{The violation of the total probability formula} \label{TPF}
This section builds on the ideas developed in \cite{soulas2024interpretation, soulas2024ontological}. According to Di Biagio and Rovelli \cite{di2021stable}, the deep difference between classical and quantum is the way probabilities behave. All classical phenomena satisfy the total probability formula (TPF)
\begin{align} \mathbb{P}(B=b) =  \sum_{a \in \mathrm{Im}(A)} \mathbb{P}(A=a) \mathbb{P}(B=b \mid A=a), \label{total_proba_formula} \end{align}
but this does not hold in general for quantum systems. As is well-know, the diagonal elements of the density matrix (when written in the eigenbasis of the observable on which the probabilities are conditioned) account for the TPF, while the non-diagonal terms are responsible for the depart from \eqref{total_proba_formula}. In short, this is because the probability to obtain an outcome $B=b$ while being in state $\rho$ is: \[ \tr(\rho \ket{b}\bra{b}) = \sum_{a, a'} \rho_{a a'} \braket{a' \vert b}  \braket{b \vert a} = \sum_{a} \underbrace{\rho_{aa}  \lvert \braket{b \vert a} \rvert^2}_{\mathbb{P}(a) \mathbb{P}(b \mid  a)}  +  \sum_{ a < a'}  \underbrace{2 \mathrm{Re}( \rho_{a a'}  \braket{a' \vert b} \braket{b \vert a })}_{\text{interferences}}. \]
Since quantum coherence is also identified with $\rho$ having non-vanishing off-diagonal elements, we propose to adopt the following guiding principle: \textit{the key property that a measure of coherence should satisfy is the ability to bound the deviation from the TPF}. 

An asset of this understanding of quantum coherence lies in its physical and ontological meaning. The fact that probabilities do not fundamentally obey \eqref{total_proba_formula} in our universe\footnote{Quantum mechanics has first been formulated in terms of non-commuting operators in a Hilbert space possibly because it offers the simplest probabilistic structure violating the TPF. Feynman's sum-over-histories approach is another natural candidate of a probabilistic framework in which all potentialities contribute.} is an experimental datum, confirmed by all kinds of interferometry experiments. This observation entails a huge philosophical consequence, because the TPF is at the core of our most intuitive notion of an objective world. Indeed, the following assertion: 
\paragraph{(P)} Even though the actual value of a variable is not known, it still has a definite value among the possible ones \\ \\
\textit{implies} the TPF.

To see this, consider a situation in which (P) holds, involving two variables, one observed $B$ and one unknown $A$ (for instance, $A \in \{ \text{Yes} , \text{No} \}$ could be the the answer to the question ‘Has my student learnt her lesson?’, and $B$ her mark obtained at the exam). Suppose that the situation has been repeated $N$ times and that the value $B=b$ has been obtained $n_b$ times. According to (P), one can discriminate between the possible cases for $A$, even without having access to it. Let's say, then, that the event $A=a$ has occured $m_a$ times, and that among these latter the value $B=b$ has been obtained $k_{a,b}$ times. One can thus write:
\[ n_b = \sum_{a} k_{a,b} =  \sum_{a} m_a \frac{k_{a,b}}{m_a}. \] 
Dividing by $N$ and taking $N \rightarrow \infty$ yields \eqref{total_proba_formula}, when the probabilities are understood as limits of empirical frequencies.

Said differently, the validity of the TPF is a necessary condition to have (P), that is to build a picture of an objective world independent of what is known about it: an ontology distinct from epistemology. The absence of such an objective ontology for quantum systems captures (one of) the core weirdness(es) of quantum theory, and can even serve as an interpretation-independent formulation of the measurement problem \cite{soulas2024interpretation, soulas2024ontological}. Decoherence can be understood as the process of objectification in some pointer basis, after which the TPF and (P) approximately hold when conditioning on observables diagonal in this preferred-basis.

To further motivate the importance of the TPF, note that it is ubiquitous in the discussions on the interpretation of quantum mechanics, although often unnoticed. The very definition of a hidden variable is an unknown quantity $\lambda \in \Lambda$ having probability distribution $\mu(\mathrm{d}\lambda)$, with respect to which any quantum observable $A$ can be conditioned so as to satisfy the TPF:
\[ \mathbb{P}(A=a) = \int_\Lambda \mathbb{P}(A =a \vert \lambda) \mu(\mathrm{d}\lambda), \]
where $\mathbb{P}(A=a \vert \lambda) = \delta(a=f(\lambda))$ is a Dirac distribution (or is at least less indeterministic than $\mathbb{P}(A=a)$), meaning that $\lambda$ can be thought of as an objective property of the system containing information about the values taken by the observables. This is already present in Bell's seminal papers \cite[equation (2)]{bell1964einstein} \cite[equation (12)]{bell1981bertlmann}, as well as in the more modern formulation of Bell's theorem in terms of causal models \cite[equation (13)]{wood2015lesson}. The assumption of ‘macroscopic realism’ used to derive the Leggett-Garg inequalities translates mathematically as a TPF as well \cite[equation (1)]{leggett1985quantum}, and it is also the basis of the ‘ontological models’ program \cite[Definition 1]{harrigan2010einstein}  \cite[equation (7)]{leifer2014quantum}. Finally, the much debated hypothesis of ‘absoluteness of observed events’, at the core of the local friendliness no-go theorem \cite{bong2020strong}, is (among other things) the assumption a probability distribution satisfying a TPF when conditioning on observed events.

We can now embark on the mathematical formalization of these ideas.

\section{A distance between orthonormal bases} \label{distance}
Following the guiding principle presented in Section \ref{TPF}, a proper measure of coherence should be a map
\[ \begin{array}{lccl} \eta : &\mathcal{S}(\mathcal{H}) \times \mathfrak{B}(\mathcal{H})  &  \longrightarrow &   \mathbb{R}_+ \\
                                                &(\rho , \mathcal{B}) & \longmapsto  & \eta_\mathcal{B}(\rho) \end{array} \]
such that for all $\rho \in \mathcal{S}(\mathcal{H})$ and all linear subspaces $F \subset \mathcal{H}$,
 \begin{equation} \lvert \underbrace{ \tr(\rho \Pi_F)}_{\substack{\text{quantum} \\ \text{probability}}}  - \;  \underbrace{\tr( D_{\mathcal{B}} \left[ \rho \right] \Pi_F)}_{\text{TPF}} \rvert \leq \dim(F) \; \eta_\mathcal{B}(\rho),   \label{bound_TPF} \end{equation}
meaning that for all probabilistic events (which, in quantum theory, correspond to subspaces), $\eta_\mathcal{B}(\rho)$ bounds the level of quantum interferences \textit{i.e.} the deviation from the TPF. Said differently, it quantifies the obstruction to being able to assign an objective value, albeit unknown, to all the observables diagonal in $\mathcal{B}$. 

Additionally, we surely want to require that the level of coherence of $\rho$ in its eigenbasis $\mathcal{B}_{\rho}$ be 0, since the system always satisfies the TPF with respect to this basis. 

Taken as such, however, these two properties constitute a very weak statement: any map such that $\eta_\mathcal{B} \geq 1$ for all $\mathcal{B} \neq \mathcal{B}_{\rho}$ would work. To make the latter more relevant, we will turn it into a continuity condition, and demand that $\eta_{\mathcal{B}} \rightarrow 0$ whenever $\mathcal{B}$ ‘tends’ to $\mathcal{B}_{\rho}$ in a sense to be now specified. To do so, we need an appropriate notion of distance between orthonormal bases.

Arguably, the distance in $\mathfrak{B}(\mathcal{H})$ best-suited for quantum mechanics is the one developed in \cite{bengtsson2007mutually}  (in the context of the search for mutually unbiased bases \cite{durt2010mutually}). It has a deep geometrical meaning, and can be written in the following simple form: if $\mathcal{B}_i = (\ket{e_i})_i$ and $\mathcal{B}_j = (\ket{f_j})_j$,
\[ d(\mathcal{B}_i , \mathcal{B}_j)^2 = \sum_{i,j} \lvert \braket{e_i \lvert f_j}  \rvert^2 \left(1 - \lvert \braket{e_i \vert f_j} \rvert^2 \right). \]
It is clear in this expression that the distance is 0 if and only if $\mathcal{B}_i$ is just a relabelling of $\mathcal{B}_j$. One can also show that, in dimension $n$, the distance lies in $[0 , \sqrt{n-1}]$ and is maximal if and only if the two bases are mutually unbiased (\textit{i.e.} such that $\forall i,j, \lvert \braket{e_i \vert f_j} \rvert^2 = \frac{1}{n}$), as are for instance the eigenbases of the spin operators $\hat{S}_X$, $\hat{S}_Y$ and $\hat{S}_Z$ (the eigenbases of $\hat{X}$ and $\hat{P}$ can also be seen as a continuous limit of mutually unbiased bases \cite{durt2010mutually}). 

It is useful to get some intuition on the behaviour of this distance. We expect that two observables with close eigenbases should almost commute and, conversely, that two almost-commuting observables should have close eigenbases. This is true, according to the following inequalities.

\begin{proposition} \label{inequalities}
Let $\hat{A}$ and $\hat{B}$ be two Hermitian operators in finite dimension $n$ with respective spectra $(a_i)_{1\leq i \leq n}$ and $(b_j)_{1\leq j \leq n}$. Then:
\[ \triplenorm{ [\hat{A},\hat{B}] } \leq \frac{\sqrt{n}}{2} C_{\hat{A}, \hat{B}} \;  d(\mathcal{B}_{\hat{A}} , \mathcal{B}_{\hat{B}}),\]
where $C_{\hat{A}, \hat{B}} = \underset{i , j} \max \lvert a_i - a_j \rvert \; \underset{k , l} \max \lvert b_k - b_l \rvert$. If moreover their spectra are non-degenerate, we have:
\[ d(\mathcal{B}_{\hat{A}} , \mathcal{B}_{\hat{B}}) \leq \frac{\sqrt{2n}}{c_{\hat{A}, \hat{B}}} \triplenorm{ [\hat{A},\hat{B}] },\]
where $c_{\hat{A}, \hat{B}} = \underset{i \neq j} \min \lvert a_i - a_j \rvert \; \underset{k \neq l} \min \lvert b_k - b_l \rvert$.
\end{proposition}

To derive these inequalities, we start with a preliminary lemma.

\begin{lemma}[Almost equality in quadratic Jensen's inequality]
Let $(x_i)_{1\leq i \leq n}$ be a collection of $n$ distinct real numbers, and $(\lambda_i)_{1\leq i \leq n} \in \mathbb{R}_+^n$ be non-negative weights such that $\sum_i \lambda_i =1$. By Jensen's inequality, we know that $\sum_i \lambda_i x_i^2 \geq (\sum_i \lambda_i x_i)^2$. If the bound is almost tight, \textit{i.e.} if $\sum_i \lambda_i x_i^2 - (\sum_i \lambda_i x_i)^2 \leq \varepsilon$, then:
\[ \sum_i \lambda_i (1-\lambda_i) \leq \frac{2 \varepsilon}{ \underset{i \neq j} \min \lvert x_i - x_j \rvert^2}. \]
\end{lemma}

\begin{proof}[Proof of Lemma]
Define $\Delta_{ij} = x_i - x_j \neq 0$, and observe that:
\begin{align*}
\sum_i \lambda_i x_i^2 - \left( \sum_i \lambda_i x_i \right)^2 
&= \sum_i \lambda_i \underbrace{(1-\lambda_i)}_{= \sum_{j \neq i} \lambda_j} x_i^2 - \sum_{i \neq j} \lambda_i \lambda_j x_i x_j \\
&= \sum_i \lambda_i x_i \sum_{j \neq i} \lambda_j (x_i - x_j) \\
&= \sum_{i \neq j} \lambda_i \lambda_j x_i \Delta_{ij} \\
&= \sum_{i < j} \lambda_i \lambda_j x_i \Delta_{ij} + \sum_{i > j} \lambda_i \lambda_j x_i \Delta_{ij} \\
&= \sum_{i < j} \lambda_i \lambda_j x_i \Delta_{ij} - \sum_{i < j} \lambda_i \lambda_j x_j \Delta_{ij}  \quad \quad \text{(relabelling the second sum)} \\
&= \sum_{i < j} \lambda_i \lambda_j \Delta_{ij}^2.
\end{align*}
Now, because all the terms are non-negative,
\begin{align*}
\sum_{i < j} \lambda_i \lambda_j \Delta_{ij}^2 \leq \varepsilon 
&\Rightarrow \sum_{i < j} \lambda_i \lambda_j \leq \frac{\varepsilon}{\underset{i \neq j} \min \Delta_{ij}^2} \\
&\Rightarrow \sum_i \lambda_i (1-\lambda_i) = \sum_{i \neq j} \lambda_i \lambda_j \leq \frac{2\varepsilon}{\underset{i \neq j} \min \Delta_{ij}^2}.
\end{align*}
\end{proof}

\begin{proof}[Proof of Proposition \ref{inequalities}]
\textit{First inequality.} Denote $\mathcal{B}_{\hat{A}} = (\ket{e_i})_{1\leq i \leq n}$ and $\mathcal{B}_{\hat{B}} = (\ket{f_j})_{1\leq j \leq n}$ the eigenbases of $\hat{A}$ and $\hat{B}$ properly associated with the eigenvalues $(a_i)_{1\leq i \leq n}$ and $(b_j)_{1\leq j \leq n}$, so that in particular $\hat{B} = \sum_{j} b_j \ket{f_j} \bra{f_j}$, and let $\tilde{C}_{\hat{A}, \hat{B}} = \underset{i , j} \max \lvert a_i - a_j \rvert \; \underset{k} \max \lvert b_k \rvert$. For all $k \in \llbracket 1, n \rrbracket$, 
\[ \hat{B} \ket{e_k} =  \sum_{i=1}^n \braket{e_i \vert \hat{B} e_k} \ket{e_i} = \sum_{1 \leq i,j \leq n} b_j \braket{e_i \vert f_j} \braket{f_j \vert e_k} \ket{e_i}. \]

Hence: 
\begin{align*}
\left\lVert (\hat{A} \hat{B} - \hat{B} \hat{A})\ket{e_k} \right\rVert^2 
&= \left\lVert \sum_{1 \leq i,j \leq n} (a_i -a_k) b_j\braket{e_i \vert f_j} \braket{f_j \vert e_k} \ket{e_i} \right\rVert^2 \\
&= \sum_{\substack{i=1 \\ i \neq k}}^n \left\lvert (a_i - a_k) \sum_{j=1}^n  b_j \braket{e_i \vert f_j}\braket{f_j \vert e_k} \right\rvert^2 \\
&\leq \sum_{\substack{i=1 \\ i \neq k}}^n \lvert a_i - a_k \rvert^2 \left( \sum_{j=1}^n \lvert b_j \rvert \lvert \braket{e_i \vert f_j} \braket{f_j \vert e_k} \rvert \right)^2 \\
&\underset{\text{C.-S.}} \leq \tilde{C}_{\hat{A}, \hat{B}}^2 \sum_{\substack{i=1 \\ i \neq k}}^n \sum_{j=1}^n \lvert \braket{e_i \vert f_j} \rvert^2 \lvert \braket{e_k \vert f_j} \rvert^2\sum_{l=1}^n 1^2 \\
&\leq n \tilde{C}_{\hat{A}, \hat{B}}^2 \sum_{j=1}^n  \lvert \braket{e_k \vert f_j} \rvert^2   \sum_{\substack{i=1 \\ i \neq k}}^n \lvert \braket{e_i \vert f_j} \rvert^2 \\
&\leq n \tilde{C}_{\hat{A}, \hat{B}}^2  \sum_{j=1}^n \lvert \braket{e_k \vert f_j} \rvert^2 (1 - \lvert \braket{e_k \vert f_j} \rvert^2).
\end{align*}
Consequently, for all $\ket{\Psi} = \sum_{k=1}^n \alpha_k \ket{e_k}$ of norm 1, 
\[ \left\lVert [\hat{A},\hat{B}]\ket{\Psi} \right\rVert^2 \leq \sum_{k=1}^n \underbrace{\lvert \alpha_k \rvert^2}_{\leq 1}  \left\lVert [\hat{A},\hat{B}]\ket{e_k} \right\rVert^2 \leq n \tilde{C}_{\hat{A}, \hat{B}}^2 \sum_{1 \leq j,k \leq n} \lvert \braket{e_k \vert f_j} \rvert^2 (1 - \lvert \braket{e_k \vert f_j} \rvert^2) =  n \tilde{C}_{\hat{A}, \hat{B}}^2 d(\mathcal{B}_{\hat{A}} , \mathcal{B}_{\hat{B}})^2. \]
Taking the supremum over $\ket{\Psi} \in \mathbb{S}^{n}$ and a square root yields the first inequality with $\tilde{C}$ instead of $C$ and $\sqrt{n}$ instead of $\frac{\sqrt{n}}{2}$. To improve the constant, notice that $\triplenorm{ [\hat{A},\hat{B}] }$ and $d(\mathcal{B}_{\hat{A}} , \mathcal{B}_{\hat{B}})$ are unchanged if one replaces $\hat{B}$ by $\hat{B} - \lambda \bbone$. Hence, for all $\lambda \in \mathbb{R}$, 
\[ \triplenorm{ [\hat{A},\hat{B}] } \leq \sqrt{n}  \; \underset{i , j} \max \lvert a_i - a_j \rvert \; \underset{k} \max \lvert b_k - \lambda \rvert  \; d(\mathcal{B}_{\hat{A}} , \mathcal{B}_{\hat{B}}). \]
The optimal choice of $\lambda$ corresponds to having $\min \mathrm{spec}(\hat{B}-\lambda \bbone) = - \max \mathrm{spec}(\hat{B}-\lambda \bbone)$, in which case $\underset{k} \max \lvert b_k - \lambda \rvert = \frac{1}{2} \underset{k , l} \max \lvert b_k - b_l \rvert$, yielding the announced result.

\textit{Second inequality.} We now suppose that the $(a_i)_{1\leq i \leq n}$ are distincts, as well as the $(b_j)_{1\leq j \leq n}$. Based on the computations above, we know that for all $k \in \llbracket 1, n \rrbracket$,
\begin{align*}
& \left\lVert (\hat{A} \hat{B} - \hat{B} \hat{A})\ket{e_k} \right\rVert^2 = \sum_{\substack{i=1 \\ i \neq k}}^n \left\lvert (a_i - a_k) \braket{e_i \vert \hat{B} e_k} \right\rvert^2\\
\Rightarrow \quad &\sum_{\substack{i=1 \\ i \neq k}}^n \left\lvert \braket{e_i \vert \hat{B} e_k} \right\rvert^2 \leq \frac{ \triplenorm{ [\hat{A},\hat{B}]}^2}{\underset{i \neq j} \min \lvert a_i - a_j \rvert^2}.
\end{align*}
Furthermore, recalling that $\braket{e_k \vert \hat{B} e_k} \in \mathbb{R}$,
\begin{align*}
\sum_{\substack{i=1 \\ i \neq k}}^n \left\lvert \braket{e_i \vert \hat{B} e_k} \right\rvert^2 
&= \left\lVert \hat{B} \ket{e_k} \right\rVert^2 -\braket{e_k \vert \hat{B} e_k}^2 \\
&= \left( \sum_{i=1}^n b_i \braket{e_k \vert f_i} \bra{f_i} \right) \cdot \left( \sum_{j=1}^n b_j \ket{f_j} \braket{f_j \vert e_k} \right) - \left( \sum_{i=1}^n b_i \lvert \braket{f_i \vert e_k} \rvert^2 \right)^2 \\
&= \sum_{i=1}^n b_i^2  \lvert \braket{e_k \vert f_i}  \rvert^2 - \left( \sum_{i=1}^n b_i \lvert \braket{e_k \vert f_i} \rvert^2 \right)^2.
\end{align*}
From this, the lemma immediately implies that:
\[ \sum_{i=1}^n \lvert \braket{e_k \vert f_i}  \rvert^2 (1 - \lvert \braket{e_k \vert f_i} \rvert^2) \leq \frac{2\triplenorm{ [\hat{A},\hat{B}]}^2}{c_{\hat{A}, \hat{B}}^2}. \]
Summing over all $k$ and taking a square root yields the second inequality.
\end{proof}

\section{Axioms for measures of coherence} \label{axioms}
We are now in position to formulate an axiomatic definition of a measure of coherence, recalling the intuition of \eqref{bound_TPF}.

\begin{definition}[Measure of coherence] \label{definition}
A map \[ \begin{array}{lccl} \eta : &\mathcal{S}(\mathcal{H}) \times \mathfrak{B}(\mathcal{H})  &  \longrightarrow &   \mathbb{R}_+ \\
                                                &(\rho , \mathcal{B}) & \longmapsto  & \eta_\mathcal{B}(\rho) \end{array} \]
is called a \textit{measure of coherence} if it satisfies the following properties 1 and 2 below.
\begin{enumerate}
\item $\underset{\mathcal{B} \overset{d} \rightarrow \mathcal{B}_{\rho}} \lim \eta_{\mathcal{B}}(\rho) = \eta_{\mathcal{B}_\rho}(\rho) = 0$;
\item for all $\rho \in \mathcal{S}(\mathcal{H})$ and all subspaces $F \subset \mathcal{H}$,
 \[ \left\lvert \tr(\rho \Pi_F)  - \; \tr( D_{\mathcal{B}} \left[ \rho \right] \Pi_F) \right\rvert \leq \dim(F) \; \eta_\mathcal{B}(\rho). \]
\end{enumerate}
\end{definition}

The limit in (1) is to be understood in the sense of the distance $d$ defined in the previous Section. Note that it is pointless to require $\eta_{\mathcal{B}}(\rho) = 0 \Leftrightarrow \mathcal{B} = \mathcal{B}_{\rho}$ because the direct implication stems from 2. Our main result is the following theorem. Here, $\left\lVert Q_{\mathcal{B}} \left[ \rho \right]  \right\rVert_p$ stands for the $p$-norm of the non-diagonal elements of $\rho$ (seen as a $n^2$-uple) as written in $\mathcal{B}$.

\begin{theorem} \label{main}
Denote $\mathcal{B} = (\ket{e_i})_{1 \leq i \leq n}$. The following maps:
\begin{align*}
\bullet \quad &\eta_1 : (\rho , \mathcal{B}) \longmapsto \left\lVert Q_{\mathcal{B}} \left[ \rho \right] \right\rVert_1 = \sum_{i \neq j} \lvert \braket{e_i \vert \rho e_j } \rvert \\
\bullet \quad &\eta_2 : (\rho , \mathcal{B}) \longmapsto \left\lVert Q_{\mathcal{B}} \left[ \rho \right] \right\rVert_2 = \left( \sum_{i \neq j} \lvert \braket{e_i \vert \rho e_j } \rvert^2 \right)^{1/2} \\
\bullet \quad &\eta_\infty :  (\rho , \mathcal{B}) \longmapsto n \left\lVert Q_{\mathcal{B}} \left[ \rho \right] \right\rVert_\infty = n \; \underset{i \neq j} \max  \lvert \braket{e_i \vert \rho e_j } \rvert \\
\bullet \quad &\delta : (\rho , \mathcal{B}) \longmapsto d(\mathcal{B}_{\rho}, \mathcal{B})
\end{align*}
are measures of coherence. On the other hand, for any choice of the constant $c>0$, the relative entropy of coherence \[ S_{\text{rel}} : (\rho , \mathcal{B}) \longmapsto c \left[ S\left( D_{\mathcal{B}} [ \rho] \right) - S(\rho) \right] \] is never a measure of coherence. 
\end{theorem}

\begin{proof}
\textit{Property 1.} Clearly, for $p=1$, $p=2$ or $p=\infty$, $\eta_p(\rho , \mathcal{B}_{\rho}) = 0$ since $\left\lVert 0 \right\rVert_p = 0$; and $d(\mathcal{B}_{\rho}, \mathcal{B}_{\rho}) = 0$. In addition, diagonalizing $\rho = \sum_k \lambda_k \ket{f_k} \bra{f_k}$ in its eigenbasis $\mathcal{B}_{\rho} = (\ket{f_k})_{1 \leq k \leq d}$ yields:

\begin{align*}
\eta_2 (\rho , \mathcal{B})^2 &= \sum_{i \neq j} \left\lvert \bra{e_i} \left( \sum_k \lambda_k \ket{f_k} \bra{f_k} \right)  \ket{e_j} \right\rvert^2 \\
&\underset{\text{Jensen}} \leq  \sum_{i,k}  \underbrace{ \lambda_k }_{\in [0,1]} \lvert \braket{e_i \vert f_k} \rvert^2  \sum_{j \neq i}  \lvert \braket{f_k \vert e_j} \rvert^2 \\
&\leq \sum_{i,k} \lvert \braket{e_i \vert f_k} \rvert^2  (1 - \lvert \braket{e_i \vert f_k} \rvert^2) = d(\mathcal{B}_{\rho}, \mathcal{B})^2  \underset{\mathcal{B} \overset{d} \rightarrow \mathcal{B}_{\rho}} \longrightarrow 0.
\end{align*}
which ensures property 1 for $\eta_2$ and $\delta$. Since $\eta_1 (\rho , \mathcal{B}) \leq n \eta_2 (\rho , \mathcal{B})$ (using Cauchy-Schwarz) and $\eta_\infty (\rho , \mathcal{B}) \leq n \eta_2 (\rho , \mathcal{B})$, property 1 is actually verified for all of them.

\textit{Property 2.} Since we have just proved that $\eta_2 \leq \delta$, it suffices to show property 2 for $\eta_1$, $\eta_2$ and $\eta_\infty$. Let's first show that $\triplenorm{ Q_{\mathcal{B}} \left[ \rho \right] } \leq \eta_p(\rho , \mathcal{B})$ for $p=2$ and $p=\infty$. This is true, because for all vectors $\ket{\Psi} = \sum_k \alpha_k \ket{e_k} \in \mathcal{H}$ of norm 1, 
\begin{align*}
\lVert Q_{\mathcal{B}} \left[ \rho \right] \ket{\Psi} \rVert^2 &= \left\lVert \sum_{i \neq j} \braket{e_i \vert \rho e_j}  \alpha_j \ket{e_i} \right\rVert^2 \\
&= \sum_i \Big\lvert \sum_{j \neq i} \braket{e_i \vert \rho e_j} \alpha_j \Big\rvert^2 \\
&\leq \sum_i \left(\sum_{j \neq i} \lvert  \braket{e_i \vert \rho e_j} \rvert \lvert \alpha_j \rvert \right)^2 \\
& \underset{\text{C.-S.}} \leq \sum_i  \sum_{j \neq i} \lvert \braket{e_i \vert \rho e_j}  \rvert^2  \underbrace{\sum_{k \neq i} \lvert \alpha_k \rvert^2}_{\leq 1} \\
&\leq \eta_2(\rho , \mathcal{B})^2.
\end{align*}
And similarly: 
\begin{align*}
\lVert Q_{\mathcal{B}} \left[ \rho \right] \ket{\Psi} \rVert^2 \leq \sum_i  \sum_{j \neq i} \lvert \braket{e_i \vert \rho e_j}  \rvert^2 \leq \frac{1}{n^2} \eta_\infty(\rho , \mathcal{B})^2 \sum_i  \sum_{j \neq i} 1 \leq \eta_\infty(\rho , \mathcal{B})^2.
\end{align*}

Now, if $F$ is a subspace of $\mathcal{H}$, let $(\ket{\pi_k})_k$ be an orthonormal basis of $F$ and $\Pi_F$ the orthogonal projector onto $F$. Then for all $\rho \in \mathcal{S}(\mathcal{H})$:
\begin{align*}
& \tr(\rho \Pi_F) -  \tr( D_{\mathcal{B}} [\rho] \Pi_F) = \tr( Q_{\mathcal{B}} [ \rho ] \Pi_F)  = \sum_{k=1}^{\dim(F)} \braket{\pi_k \vert Q_{\mathcal{B}} [ \rho ] \pi_k} \\
\Rightarrow \quad & \lvert \tr(\rho \Pi_F) -  \tr( D_{\mathcal{B}} [ \rho ] \Pi_F) \rvert \leq \sum_{k=1}^{\dim(F)} \triplenorm{ Q_{\mathcal{B}} [ \rho ] } \leq \dim(F) \eta_p(\rho , \mathcal{B}),
\end{align*} 
for $p=2$ or $p=\infty$. Finally, for $p=1$, remark that:
\[  \lvert \tr(\rho \Pi_F) -  \tr( D_{\mathcal{B}} [ \rho ] \Pi_F) \rvert \leq \sum_{k=1}^{\dim(F)} \lvert \braket{\pi_k \vert Q_{\mathcal{B}} [ \rho ] \pi_k} \rvert \leq \sum_{k=1}^{\dim(F)} \sum_{i\neq j} \lvert \braket{\pi_k \vert e_i} \braket{e_i \vert \rho e_j} \braket{e_j \vert \pi_k} \rvert \leq \dim(F) \eta_1(\rho , \mathcal{B}). \]

It remains to find a counter-example for $S_{\text{rel}}$. For some $\varepsilon \in [0,1]$, consider for instance $\rho = \begin{pmatrix}  \frac{1}{2} & \frac{1}{2}\varepsilon \\  \frac{1}{2}\varepsilon & \frac{1}{2} \end{pmatrix}$ as written in $\mathcal{B}$ and $F = \mathbb{C} \cdot \left( \frac{\ket{e_1} + \ket{e_2} }{\sqrt{2}} \right)$ of dimension 1, whose associate orthogonal projector is $\Pi_F = \begin{pmatrix}  \frac{1}{2} & \frac{1}{2} \\  \frac{1}{2} & \frac{1}{2} \end{pmatrix}$. Then $\lvert \tr(\rho \Pi_F) -  \tr( D_{\mathcal{B}} [ \rho ] \Pi_F) \rvert = \frac{\varepsilon}{2}$ but $S_{\text{rel}} (\rho , \mathcal{B}) = c \left( \ln(2) + \frac{1+\varepsilon}{2} \ln\left(  \frac{1+\varepsilon}{2} \right) + \frac{1-\varepsilon}{2} \ln\left[ \frac{1- \varepsilon}{2} \right) \right] = \frac{c}{2}  \varepsilon^2 + o(\varepsilon^2)$, therefore one can always find an $\varepsilon$ such that property 2 is not satisfied.
\end{proof}

The measure $\eta_\infty$ (or rather $\frac{1}{n} \eta_\infty$), which uniformly bounds the off-diagonal elements, is widely used in the literature on decoherence, although often implicitly and somewhat informally. Depending on the context, it can be assimilated with the decoherence factor (among numerous examples, see for instance \cite{kiefer2000decoherence, riedel2012rise}) or with the typical time of decoherence via its logarithm. Up to a slight modification, it corresponds to the quantity $\varepsilon$ introduced in \cite{di2021stable} and studied in \cite{soulas2024decoherence}.

Also well-known are the measures $\eta_1$ and $\eta_2$, generally called $l^1$ and $l^2$-norms of coherence \cite{streltsov2017colloquium}. Note that, according to the definition proposed in \cite{baumgratz2014quantifying}, $\eta_2$ is \textit{not} a measure of coherence (nor is $\eta_\infty$: the counter-example given in \cite[Appendix B]{baumgratz2014quantifying} also leads to a contradiction with $\eta_\infty$ for $\lvert \beta \rvert > \frac{1}{6}$), whereas the relative entropy of coherence is one. Therefore, Definition \ref{definition} indeed captures a different class of measures of coherence than the one of Baumgratz, Cramer, and Plenio, which is neither strictly smaller, nor strictly larger.

Finally, $\delta$ is probably a new measure that hasn't been proposed in the past, although not a very interesting one, because we always have $\delta \geq \eta_2$.

\section{Purity as the approximate level of $l^2$-coherence in most bases} \label{l2}
As noted in the introduction, the framework of resource theory mainly used so far to quantify coherence do not naturally invite to study the dependence of coherence on the basis. In this section, we present such a result concerning $\eta_2$, which allows to give a physical meaning to (the square root of) the purity of a state: $\sqrt{\tr(\rho^2)}$ \textit{can be seen as the approximate level of $l^2$-coherence in most bases}.

\begin{proposition} \label{purity}
Let $\mathcal{H}$ be an $n$-dimensional Hilbert space, $\rho \in \mathcal{S}(\mathcal{H})$ and $\mathcal{B}$ be a random orthonormal basis of $\mathcal{H}$, defined as $\mathcal{B} = (U\ket{e_i})_{1\leq i \leq n}$ where $(\ket{e_i})_{1\leq i \leq n}$ is a reference orthonormal basis and $U$ a random unitary following the Haar measure on $\mathcal{U}(\mathcal{H})$. Then the $l^2$-norm of coherence in $\mathcal{B}$ is a random variable satisfying:
\[ \mathbb{E}\left(\lvert \eta_2(\mathcal{B}, \rho)^2 -  \tr(\rho^2) \rvert\right)  \underset{n \rightarrow +\infty} \longrightarrow 0. \]
\end{proposition} 
For the sake of rigour, we do not write $\eta_2(\mathcal{B}, \rho)^2 \underset{n \rightarrow \infty}{\overset{\text{‘}L^1\text{’}}{ \xrightarrow{\hspace*{0.8cm}}}}   \tr(\rho^2)$ because the space $\mathcal{H}$ on which the randomness is defined changes with $n$.

\begin{proof}
Let $\mathcal{B}_\rho = (\ket{e_i})_{1\leq i \leq n}$ be $\rho$'s eigenbasis, in which $\rho =  \sum_k \lambda_k \ket{e_k}\bra{e_k}$ and $\eta_2(\mathcal{B}_\rho, \rho) = 0$. Without loss of generality, we can take $\mathcal{B}_\rho$ as the reference basis of the proposition, since the Haar measure is invariant under composition with a fixed unitary. Expressed in $\mathcal{B}$, the i$^{th}$ diagonal coefficient of $\rho$ is: $\rho_{ii} = \braket{ U e_i \vert \rho \vert U e_i } = \braket{Ue_i \vert \sum_k \lambda_k \ket{e_k}\bra{e_k} U e_i} = \sum_k \lambda_k \lvert \braket{e_i \vert U e_k} \rvert^2$. As $U$ follows a uniform law on $\mathcal{U}(\mathcal{H})$, so does $U\ket{e_k}$ on the complex $n$-sphere $\mathbb{S}^{n} \subset \mathcal{H}$. Define the following random variables:  $X_{ik} = \lvert \braket{e_i \vert U e_k} \rvert^2$, $Y_i = \sum_k \lambda_k X_{ik} =\rho_{ii}$, and $T = \sum_i Y_i^2 =  \sum_i \rho_{ii}^2$. 

To prove the proposition, it suffices to show that $\mathbb{E}(T) \underset{d \rightarrow +\infty} \longrightarrow 0$. Indeed,
\[ \eta_2(\mathcal{B}, \rho)^2 = \sum_{i \neq j} \lvert \rho_{ij} \rvert^2 = \tr(\rho^2) - \sum_i \rho_{ii}^2.\]

An efficient way to achieve this is to use the following formula \cite[Theorem 34]{kostenberger2021weingarten} \cite[Theorem 29.9]{hewitt2013abstract}: for $\textbf{a} = (a_1, \dots, a_n)$ such that $a_1 + \dots + a_n = m$, we have
\[  \int_{\mathcal{U}(n)} \prod_{k=1}^n \lvert u_{1k} \rvert^{2a_k} \mathrm{d}\mu_n = \frac{(n-1)!}{(m+n-1)!} \prod_{k=1}^n a_k !, \]
where $\mathrm{d}\mu_n$ is the Haar measure on the unitary group in dimension $n$ and $u_{ik} = \braket{e_i \vert U e_k}$ is the $(i,k)$ coefficient of the unitary matrix $U$. Of course, one can replace the line index 1 by any index $i$, since the Haar measure remains unchanged when composed with a given permutation matrix. Applied to $m=2$ and $\textbf{a} = (2, 0 \dots, 0)$ or $\textbf{a} = (1, 1 \dots, 0)$, we get:
\begin{align*}
\mathbb{E}\left( X_{ik} X_{il} \right) &=  \int_{\mathcal{U}(\mathcal{H})} \lvert u_{ik} \rvert^{2} \lvert u_{il} \rvert^{2} \mathrm{d}\mu_n
&=  \left\{ \begin{array}{ll} \frac{2(n-1)!}{(2+n-1)!} = \frac{2}{n(n+1)} \text{ if } k=l \\ \\  \frac{(n-1)!}{(2+n-1)!} = \frac{1}{n(n+1)} \text{ if } k\neq l \end{array} \right. 
&= \frac{1}{n(n+1)} (\delta_{kl} +1).
\end{align*}
Thus:
\begin{align*}
\mathbb{E}\left( Y_i^2 \right) &= \sum_{1 \leq k,l \leq n} \lambda_k \lambda_l \mathbb{E}\left( X_{ik} X_{il} \right)
&= \frac{1}{n(n+1)} \left( \sum_{k=1}^n \lambda_k^2 + \sum_{1 \leq k,l \leq n} \lambda_k \lambda_l \right)
&=  \frac{\tr(\rho^2) +1}{n(n+1)}.
\end{align*}
Finally:
\[\mathbb{E}\left( T \right) = \sum_{i=1}^d \mathbb{E}\left( Y_i^2 \right) = \frac{\tr(\rho^2) +1}{n+1} \underset{n \rightarrow +\infty} \longrightarrow 0, \]
which completes the proof.
\end{proof}

\begin{remark}
This proposition gives the approximate level of $l^2$-coherence for any subsystem of the universe (provided it is of dimension $d\gg1$) in most bases. Said differently, when performing experiments on any (sufficiently large) system, the typical deviation from the TPF will be approximately the same for all observables, and close to $\sqrt{\tr(\rho^2)}$.

One might even suspect a link with the rise of entropy. As time flows, entropy is likely to increase with our statistical ignorance, therefore $\rho$ is expected to get closer to the maximally entangled state and the purity to $\frac{1}{n}$, which goes to 0 as $n \rightarrow +\infty$. Thus, the TPF tends to be satisfied in all bases because of the rise of our ignorance.

This feature (\textit{i.e.} the nearly constant level of coherence in most bases) is actually expected for \textit{any} sufficiently smooth measure of coherence, due to the mathematical phenomenon known as ‘concentration of measure’. Indeed, since the unitary groups equipped with their Haar measures $(\mathcal{U}(n) , \mu_n)_{n \geq 1}$ constitute a Lévy family, any Lipschitz map $f_n : \mathcal{U}(n) \rightarrow \mathbb{R}$ will have the property to be approximately equal to its median value except on a region that becomes exponentially small when $n \rightarrow +\infty$ \cite[Chapter 6]{milman1986asymptotic}. Of course, the interest of Proposition \ref{purity} is to explicitly give this median for $\eta_2$ and some bounds of convergence.
\end{remark}

\section{Conclusion} \label{conclusion}

In this paper, we have proposed a new way to think of quantum coherence as the ability for a quantum system's statistics to deviate from the total probability formula. This approach has two kinds of advantages compared to the previous proposals. First, it quantifies coherence in \textit{any} basis, rather than in a single computational basis; this allows in particular to study how coherence depends on the basis of interest (as exemplified by the result of Section \ref{l2}). Second, it has a clear physical and ontological meaning (motivated in Section \ref{TPF}), relevant in the discussions on the interpretations of quantum theory. Although we have checked whether our Definition \ref{definition} were satisfied by several natural candidates, many other common measures remain to be examined, in order to better characterize this class of measure, in particular in comparison with that of \cite{baumgratz2014quantifying}.

\section*{Acknowledgements}
I would like to gratefully thank Dimitri Petritis for the great freedom he has granted me in my research during my PhD, and to Časlav Brukner for having accepted me as a postdoctoral researcher in his group.

\section*{Declarations}
\textbf{Conflict of interest statement:} the author has no conflicts to disclose. \\
\textbf{Ethics approval, Consent, Data, Materials and code availability, Authors’ contribution statements:} not applicable.

\bibliographystyle{siam}
\bibliography{Biblio_quantifying}
\end{document}